\documentclass[]{aa}
\usepackage{times}
\usepackage{graphics}

\begin{document}

\thesaurus{08(08.19.5 SN 1993J; 10.08.1; 11.09.3; 11.09.4; 11.17.1; 13.21.5)}

\title{Absorption line systems on the line of sight from SN\,1987A to
       SN\,1993J and the intergalactic cloud in front of M\,81}

\author{Ole Marggraf
        \and
        Klaas S. de Boer
} 

\offprints{O. Marggraf (marggraf@astro.uni-bonn.de)}

\institute{Sternwarte der Universit\"at Bonn, Auf dem H\"ugel 71, 
           D-53121 Bonn, Germany
}

\date{Received 4 November, 1999 / accepted 5 September, 2000}

\titlerunning{Absorption line systems on the line of sight from SN\,1987A to
              SN\,1993J}
\authorrunning{O. Marggraf \& K.S. de Boer}
\maketitle

\begin{abstract}
We present an extended analysis of the high dispersion IUE spectra of 
\object{SN\,1993J} in \object{M\,81}, 
based on the improved data in the IUE final archive.
Eight interstellar absorption components can be identified in the velocity 
range from $-200$~km\,s$^{-1}$ to $+200$~km\,s$^{-1}$.
Column densities are determined for \ion{Fe}{ii}, \ion{Mg}{i}, \ion{Mg}{ii},
\ion{O}{i}, \ion{S}{ii}, \ion{Si}{ii}, and \ion{Zn}{ii}. 
From the ion ratios, we estimate electron densities and temperatures of the 
gas in these components.
In addition, we rederive
column density profiles for the higher ionisation stages \ion{Al}{iii}, 
\ion{Si}{iv}, and \ion{C}{iv} for the entire line of sight. 

Of special interest are the absorption components at positive velocities,
which are associated with the intergalactic medium (IGM) in the foreground of 
\object{M\,81}.
The medium has many gas components, as known from radio measurements,
probably released in dynamical interactions by galaxies in the \object{M\,81}
group.
The IUE spectra contain the only UV absorption information on such gas.
We find neutral and ionised gas absorption at the same velocity, having a 
metal content similar to a galactic disk.

We combine the spectra towards \object{SN\,1993J} in \object{M\,81} with those
towards \object{SN\,1987A} in the LMC, the directions being essentially
antipodes on the celestical sphere.
Comparing the total profile with QSO spectra, we conclude that most absorption
in QSO spectra occurs in diffuse clouds on those sight lines and is not 
similar to absorption by galactic disk gas.
However, one absorption component towards \object{SN\,1993J} is clearly very 
similar to some of the absorption components in QSO spectra.

          
\end{abstract}

%%%%%%%%%%%%%%%%%%%%%%%%%%%%%%%%%%%%%%%%%%%%%%%%%%%%%%%%%%%%%%%%%%%%%%%%%%%%%%

\section{Introduction}
%%%%%%%%%%%%%%%%%%%%%%

In recent years two supernovae occurred in nearby galaxies, \object{SN\,1987A}
in the \object{LMC} and \object{SN\,1993J} in \object{M\,81}.
High resolution UV spectra of both supernovae have been taken with the IUE 
satellite in the full wavelength range from 1150 up to 3200\AA. 
A large number of absorption lines from various metals is located in this 
part of the spectrum, 
providing valuable information on the Galactic and extragalactic interstellar 
medium along the lines of sight to the two neighbouring galaxies.

A first analysis of the ISM in the direction of \object{SN\,1993J} using IUE 
spectra was made by de Boer et al. (\cite{deboer93}).
Optical spectra of the supernova were analysed by Vladilo et al. 
(\cite{vladilo93}, \cite{vladilo94}).
High velocity clouds (HVCs) and intermediate velocity clouds (IVCs) are 
present.
These designations refer to gas with velocity not compatible with what is 
expected from Galactic rotation.
Rather arbitrarily, HVC is used for $|v_{\mathrm{rad}}|>100$~km\,s$^{-1}$ and 
IVC for $100$~km\,s$^{-1}>|v_{\mathrm{rad}}|\ga25$~km\,s$^{-1}$.
We extend the previous analysis of IUE spectra by using the 
rereduced spectra from the final archive (IUEFA\footnote{WWW URL: 
{\tt http://iuearc.vilspa.esa.es}}) 
to determine column densities for the metals in the neutral components. 
We rederive the column density profiles for the high ion stages of 
\ion{Al}{iii}, \ion{Si}{iv} and \ion{C}{iv}.

The absorption components at velocities $>50$~km\,s$^{-1}$ are of major
interest here. 
Their velocities are uncommon for the Galactic sky in this direction and the 
absorption most likely originates in or near the \object{M\,81} galaxy group
(distance of \object{M\,81} $3.63\pm0.34$~Mpc, Freedman et al. 
\cite{freedman94}).
The column densities derived are used to estimate the density and 
temperature in these clouds and to find support for the proposed location.
Appleton et al. (\cite{appleton81}) give a list of velocities for members of 
the \object{M\,81} group. 
\object{M\,81} itself has a systemic velocity of $-34$~km\,s$^{-1}$, with a 
rotational velocity of $-130$~km\,s$^{-1}$ at the position of 
\object{SN\,1993J} (Rots \cite{rots75}). 
The velocity of \object{M\,82} is $+183$~km\,s$^{-1}$; that of 
\object{NGC\,3077} $+19$~km\,s$^{-1}$.
Intergalactic molecular gas was found in CO emission southeast of the 
\object{M\,81} disk at a velocity of $-35$~km\,s$^{-1}$ (Brouillet et al. 
\cite{brouillet92}).
Also, intergalactic gas at velocities of $+120$ to $+140$~km\,s$^{-1}$
is detected in a tidal tail region connecting \object{M\,81} and 
\object{M\,82}, north of the \object{M\,81} center (Appleton et al. 
\cite{appleton81}, Yun et al. \cite{yun93}).
The gas found in the \object{SN\,1993J} absorption spectra fits into this 
range of velocities.

Combining the spectra towards \object{SN\,1993J} with those towards 
\object{SN\,1987A} (essentially on the opposite side of the sky) allows one to 
create the absorption spectrum through 3 galaxies. 
This total spectrum is compared with absorption as seen on QSO lines of sight. 
This provides a possibility to judge the origin of the absorptions.

The data used are presented in Sect.~\ref{Data}.
The results from the analysis of the \object{SN\,1993J} data are given in 
Sect.~\ref{Low_Ion} for the neutral gas and in Sect.~\ref{High_Ion} for the 
ionised gas, followed in Sect.~\ref{SN93} by a discussion of the structure 
of the \object{SN\,1993J} line of sight.

In Sect.~\ref{Long_LoS} we present the combined spectra of the full line of 
sight, followed by the comparison with QSO spectra in Sect.~\ref{QSO_LoS}. 
Finally, in Sect.~\ref{IGCs}, we summarise our understanding of the 
nature of the intergalactic clouds towards \object{SN\,1993J}
in relation to the dynamics of the \object{M\,81} group of galaxies.

%%%%%%%%%%%%%%%%%%%%%%%%%%%%%%%%%%%%%%%%%%%%%%%%%%%%%%%%%%%%%%%%%%%%%%%%%%%%%%%

\section{The data}
%%%%%%%%%%%%%%%%%%

\label{Data}

We analysed the IUE spectra provided by the ESA IUE Final Archive at VILSPA, 
Spain. 
Four spectra of \object{SN\,1993J} were obtained: two each in the short and 
long wavelength range (SWP~47386, SWP~47394, LWP~25238, LWP~25250). 
The resolution of the spectra is about $20-25$~km\,s$^{-1}$.
The dates of the spectra are listed in Table \ref{Spec_IUE}.
For a representative selection of absorption line spectra see 
Fig.~\ref{Spec_IUE}.
Due to the rapid decrease in UV flux in the supernova, the net flux
is significantly weaker for those spectra which were taken later.
This leads to a lower signal-to-noise ratio, even though these spectra have 
larger exposure times.

The \object{SN\,1993J} data from IUE can be combined with published data 
from other wavelength regions.
In particular, we use sodium column densities derived by
Vladilo et al. (\cite{vladilo93}, \cite{vladilo94}) and by Bowen et al. 
(\cite{bowen94}) from optical spectra. 
Bowen et al. (\cite{bowen94}) also derive column densities for \ion{Mg}{ii} 
from HST GHRS spectra, at a higher resolution than that of the IUE.
The Effelsberg \ion{H}{i} 21~cm spectra (Vladilo et al. \cite{vladilo94}) are
also taken into account.

A much larger number of spectra was taken for \object{SN\,1987A}.
Referring to the spectral atlas of Blades et al. (\cite{blades88a}), we use 
the short wavelength spectrum SWP~30381, which has unsaturated net flux of 
FN\,$<40\,000$ in the echelle orders $78$-$115$. 
The spectrum was taken on 25 February 1987 with an exposure time of 
$30$~min.
No long wavelength spectrum of \object{SN\,1987A} is used.

\begin{table}[t!]
\begin{flushleft}
\caption[]{IUE high dispersion spectra of \object{SN\,1993J}}
\label{Spec_IUE}
\begin{tabular}{lllr}
\hline\noalign{\smallskip}
Image     & \multicolumn{1}{c}{1993 Date} & \multicolumn{1}{c}{Time} & Exp. time [min] \\ 
\noalign{\smallskip}\hline\noalign{\smallskip}
LWP 25238 & 30 March & 06:13 UT   & $80 $\hspace{9mm} \\
SWP 47386 & 30 March & 07:58 UT   & $320$\hspace{9mm} \\
SWP 47394 & 31 March & 05:43 UT   & $655$\hspace{9mm} \\
LWP 25250 & 01 April & 03:41 UT   & $240$\hspace{9mm} \\
\noalign{\smallskip}\hline
\end{tabular}
\end{flushleft}
\end{table}

\begin{figure}[t!]
\resizebox{\hsize}{!}{\includegraphics{h1837.f1}}
\caption[]{Interstellar absorption profiles of some transitions from the 
           earlier spectra SWP~47386 and LWP~25238. 
           \ion{Zn}{ii}~(2026~\AA) blends with \ion{Mg}{i} at positive 
           velocities.
           The relative error $\sigma_n/n$ in the continuum is in the 
           vicinity of 10\%.
           The ``R'' denotes a reseau mark
          }
\label{Spectra}
\end{figure}

The spectra show the typical velocity profile seen along almost all sight 
lines towards the \object{LMC} (Savage \& de Boer \cite{savage81a}; 
de Boer \& Savage \cite{deboer80}): 
local Galactic gas around $0$~km\,s$^{-1}$, a Galactic IVC and HVC at $+60$ 
and $+125$~km\,s$^{-1}$, respectively, one absorption component of as yet 
unassigned location in front of the \object{LMC} at $+160$~km\,s$^{-1}$, 
and broad \object{LMC} absorption between $+200$ and $+330$~km\,s$^{-1}$.

Metal column densities for \object{SN\,1987A} were recently derived by 
Welty et al. (\cite{welty99}) using profile fitting. 
For the high ionisation states we use the analysis by Savage et al. 
(\cite{savage89}).
 
The \object{SN\,1993J} spectra discussed here are reduced by the NEWSIPS 
algorithms, in contrast to the analysis published by de Boer et al. 
(\cite{deboer93}).
This provides an improved background correction, especially in the higher,
closely spaced echelle orders, an increased signal-to-noise ratio, 
and a higher spectral resolution (Garhart et al. \cite{garhard97}).
We smoothed the data using a $3$ pixel wide boxcar algorithm. 
In the NEWSIPS spectra the zero intensity level of strong absorption lines
appears to lie a few percent too low. 
We have generally lifted the zero level accordingly.

%%%%%%%%%%%%%%%%%%%%%%%%%%%%%%%%%%%%%%%%%%%%%%%%%%%%%%%%%%%%%%%%%%%%%%%%%%%%%%%

\section{Lower ionisation stages in the SN\,1993J spectra}
%%%%%%%%%%%%%%%%%%%%%%%%%%%%%%%%%%%%%%%%%%%%%%%%%%%%%%%%%%

\label{Low_Ion}

\subsection{Metal and \ion{H}{i} column densities}
%=================================================

\begin{table*}[t!]
\begin{flushleft}
\caption[]{Metal column densities $\log N$ [cm$^{-2}$] towards 
           \object{SN\,1993J}}
\label{ColDensities93}
\setlength{\tabcolsep}{1.3mm}
\begin{tabular}{l*{8}{c}}
\hline\noalign{\smallskip}
    & A & B & C & D$_a$ & D$_b$ & E & F & G \\
\noalign{\smallskip}\hline\noalign{\smallskip}
$v_{\rm{LSR}}$ [km\,s$^{-1}$] & $-130$ & $-90$ & $-50$ & $-10$ & $+10$ & $+80$ & $+130$ & $+160$ \\
\noalign{\smallskip}\hline\noalign{\smallskip}
\ion{Fe}{ii} & $15.25\pm0.25$ & $13.90\pm0.30$ & $15.15\pm0.20$ & $14.95\pm0.25$ & $14.35\pm0.40$ & $14.15\pm0.25$ & $15.30\pm0.20$ & $14.15\pm0.25$ \\ 
\ion{Mg}{i}  & -- & -- & $12.65\pm0.15$ & $12.45\pm0.20$ & -- & $11.20\pm0.35$ & $12.95\pm0.35$ & $11.80\pm0.15$ \\ 
\ion{Mg}{ii} & $14.95\pm0.30$ & $>13.90$ & $14.15\pm0.15$ & $14.60\pm0.30$ & $<13.40$ & $14.70\pm0.30$ & $14.60\pm0.30$ & $14.20\pm0.30$ \\ 
\ion{O}{i}   & $14.90\pm0.30$ & $>15.85$ & -- & -- & $13.85\pm0.45$ & $15.70\pm0.50$ & $16.50\pm0.20$ & $15.25\pm0.45$ \\ 
\ion{S}{ii}  & $16.10\pm0.40$ & $14.60\pm0.20$ & $15.55\pm0.25$ & $15.65\pm0.25$ & $15.85\pm0.45$ & $15.30\pm0.40$ & $14.25\pm0.25$ & $<14.15$ \\ 
\ion{Si}{ii} & $15.05\pm0.15$ & $14.35\pm0.10$ & $15.20\pm0.10$ & $15.05\pm0.15$ & $14.45\pm0.15$ & $14.20\pm0.20$ & $14.10\pm0.20$ & $14.25\pm0.20$ \\ 
\ion{Zn}{ii} & $12.90\pm0.20$ & $12.35\pm0.25$ & $12.05\pm0.30$ & $12.95\pm0.20$ & -- & $11.85\pm0.25$ & $12.40\pm0.20$ & -- \\ 
\noalign{\smallskip}\hline\noalign{\smallskip}
$b$ [km\,s$^{-1}$]\,$^a$ & $6-7$ & $4-6$ & 9 & $6-7$ & $1.5-2$ & $5-7$ & $6-7$ & $5-6$ \\
\noalign{\smallskip}\hline\noalign{\smallskip}
\end{tabular}
$^a$ {$b$-values are derived from curves of growth.
      Their uncertainties can be up to $2$~km\,s$^{-1}$, depending on 
      the scatter of the equivalent widths}
\end{flushleft}
\end{table*}

\begin{table*}[t!]
\begin{flushleft}
\caption[]{Hydrogen column densities $\log N_{\mathrm{H}}$ [cm$^{-2}$] and 
           metal abundances 
           $\log \left(N_{\mathrm{X}}/N_{\mathrm{H}}\right)$
           derived towards \object{SN\,1993J}
           (discussion in Sect.~\ref{sub_space_highion})}
\label{Abund93}
\setlength{\tabcolsep}{1.1mm}
\begin{tabular}{l*{9}{c}}
\hline\noalign{\smallskip}
   & A & B & C & D$_a$ & D$_b$ & E & F & G & solar$^a$ \\
\noalign{\smallskip}\hline\noalign{\smallskip}
$\log N_{\rm{H}_{\rm{S,Zn}}}$ & $20.60\pm0.45$ & $19.58\pm0.32$ & $19.90\pm0.39$ & $20.40\pm0.32$ & $<21.10$ & $19.78\pm0.47$ & $19.43\pm0.32$ & $<18.95$ & \\ 
${\log N_{\rm{H}_{\rm{21\,cm}}}}^b$ & -- & -- & $19.81$ & $20.36$ & $<19.4$ & $<19.4$ & $<19.4$ & $<19.4$ & \\  
\noalign{\smallskip}\hline\noalign{\smallskip}
Fe & $-5.35\pm0.51$ & $-5.68\pm0.44$ & $-4.75\pm0.44$ & $-5.45\pm0.41$ & $>-7.15$ & $-5.63\pm0.53$ & $-4.13\pm0.38$ & $>-5.05$ & $-4.50$ \\
Mg & $-5.65\pm0.54$ & $>-5.68$ & $-5.74\pm0.42$ & $-5.80\pm0.44$ & $>-7.70$ & $-5.08\pm0.56$ & $-4.82\pm0.44$ & $>-5.05$ & $-4.46$ \\
O  & $-5.70\pm0.54$ & $>-3.73$ & -- & -- & $>-7.70$ & $-4.08\pm0.69$ & $-2,93\pm0.38$ & $>-4.15$ & $-3.16$ \\
Si & $-5.55\pm0.47$ & $-5.23\pm0.34$ & $-4.70\pm0.40$ & $-5.35\pm0.35$ & $>-6.80$ & $-5.58\pm0.51$ & $-5.33\pm0.38$ & $>-4.90$ & $-4.45$ \\
\noalign{\smallskip}\hline\noalign{\smallskip}
\end{tabular}
$^a$ solar abundances are taken from review by de Boer et al. 
  (\cite{deboer87a}) \\
$^b$ 21~cm \ion{H}{i} column densities are taken from Vladilo et al. 
(\cite{vladilo94}) \\
\end{flushleft}
\end{table*}

The wavelength scale of the IUE spectra might be affected by pointing
errors, resulting in a velocity shift for the absorption components.
We therefore fix the velocity scale by aligning, for each transition, the 
absorption at $\sim0$ and $\sim-50$~km\,s$^{-1}$ with the corresponding
clearly visible emission in the Vladilo et al. (\cite{vladilo94}) 
\ion{H}{i} 21~cm spectra.
All velocities are given with respect to the LSR.

For the identification of the absorption components we use the well exposed
transitions of \ion{Si}{ii} at 1304~\AA\, and of \ion{Fe}{ii} at 2382~\AA.
Due to the improved data reduction, we are able to resolve eight 
absorption components (see Fig. \ref{Spec_IUE})
with the velocities listed in Table \ref{ColDensities93}.
An additional component might be present at $-170$~km\,s$^{-1}$ in some 
transitions, but it is too weak for analysis.
No absorption components are seen at velocities lower than 
$-180$~km\,s$^{-1}$ or larger than $+200$~km\,s$^{-1}$. 
Due to the low signal-to-noise ratio of the spectra, the centres of the 
absorption components vary slightly and we refer to the components by rounded
radial velocities.

The component D around $0$~km\,s$^{-1}$ in many cases splits into a strong 
component D$_a$ and a weak D$_b$. 
These components, however, cannot be clearly distinguished in all transitions.
Components A and B near $-130$ and $-90$~km\,s$^{-1}$ are resolved, 
but still seem to blend in some transitions. 
We analyse them separately now, in contrast to the previous publication.
The same holds for F and the weaker component G.
The blending leads to larger uncertainties in the resulting column densities, 
especially for the weaker of these components.
Component E was disregarded in the previous publication, but is clearly
visible in most transitions.

The equivalent widths of the absorption lines are measured using a trapezium 
fit. 
Due to the faint continuum no better results would be achievable with Gaussian
fitting.  
We account for uncertainties in the continuum by determining equivalent widths
for a lowest and highest possible local continuum. 
Half of this variation is quadratically added to the error resulting from 
noise.
Metal column densities are then determined using standard curve-of-growth 
methods. 

We use the transitions of \ion{Fe}{ii} (1608, 2260, 2344, 2374, 2382, 2586, 
and 2600~\AA), \ion{Si}{ii} (1260, 1304, 1526, and 1808~\AA), \ion{Mg}{i}
(2852~\AA), \ion{Mg}{ii} (2796 and 2803~\AA), \ion{O}{i} (1302~\AA), 
\ion{S}{ii} (1250, 1253, and 1259~\AA), and of \ion{Zn}{ii} (2026~\AA); 
$f$ values are taken from the compilation by Morton (\cite{morton91}), 
plus newer values from the review by Savage \& Sembach
(\cite{savage96}).
The $b$-value for each absorption component is fixed using the equivalent 
widths of the \ion{Si}{ii} and \ion{Fe}{ii} transitions. 
The remaining equivalent widths are then fitted to the appropriate theoretical 
curve of growth.
Column densities and $b$-values are given in Table~\ref{ColDensities93}.
The errors listed represent minimum and maximum column densities possible
within the range in $b$, allowed by the errors in the equivalent widths.
All subsequent error values follow from standard error propagation.

\ion{H}{i} column densities from \ion{H}{i} 21~cm emission spectra 
(Vladilo et al. \cite{vladilo94}) can only be derived for the components C at
$-50$~km\,s$^{-1}$ and D$_a$ and D$_b$ around $0$~km\,s$^{-1}$. 
The broad emission from the disk of \object{M\,81} makes it impossible to give
reliable \ion{H}{i} column densities for the absorption components at higher
negative velocities.
At positive velocities only upper limits follow from the $21$~cm spectra.

However, we can derive \ion{H}{i} column densities for all absorption 
components from the column densities of the normally only slightly depleted 
sulfur and zinc, under the assumption of no depletion at all for these two 
metals. 
Although this assumption introduces an additional source of error, this method
provides more reliable information on \ion{H}{i} in the single components than
the 21~cm radio spectra at large cloud distances, since these are based on
integration over the full width of the radio beam.
As shown in Table \ref{Abund93}, the \ion{H}{i} column densities calculated 
this way are in good agreement with the 21~cm measurements for those 
components, for which values are given by Vladilo et al. (\cite{vladilo94}).
For the other components the column densities from the IUE data are 
equal to or larger than those from the radio emission spectra, which is
expected for clouds at a larger distance that do not fill the radio beam.
Based on these data, the metal abundances also listed in Table \ref{Abund93} 
can be calculated.
The large depletion of metals in component D$_b$ is possibly the result of
underestimated metal column densities due to blending with the strong 
component D$_a$.

\subsection{Physical conditions in the gas}
%==========================================

\label{Phys_Cond}

Important information on the physical conditions in the interstellar clouds,
such as temperatures and electron densities, can be inferred from the column 
density ratios of the atom's different ionisation stages. 
The various spectra of \object{SN\,1993J} provide column densities for two of 
these:

The ratio of $N(\mathrm{\ion{Mg}{ii}})/N(\mathrm{\ion{Mg}{i}})$ 
can be read directly from the IUE spectra for most absorption components.
If \ion{Mg}{i} column densities cannot be determined from the IUE spectra, 
we take the values of Bowen et al. (\cite{bowen94}) from GHRS spectra. 
Since these data have a higher resolution, we have to identify their
components with those found by us. 
We assume (velocities in km\,s$^{-1}$)
\begin{center}
$[-145, -115]\rightarrow$~A;\hspace{2mm}
$[-110, -70]\rightarrow$~B;\hspace{2mm} \\
$[-70, -30]\rightarrow$~C;\hspace{2mm}
$[-20, +5]\rightarrow$~D$_a$;\hspace{2mm}
$[+5, +15]\rightarrow$~D$_b$;\hspace{2mm} \\
$[+80, +105]\rightarrow$~E;\hspace{2mm}
$[+110, +150]\rightarrow$~F;\hspace{2mm}
$[+150, +175]\rightarrow$~G.
\end{center}
The resulting values are listed in Table~\ref{MgI_II}.
For \ion{Mg}{i} in D$_a$ we assume the larger column density from 
Bowen et al. (\cite{bowen94}) as an upper limit. 
The larger value in E can be due to an additional component at 
$104$~km\,s$^{-1}$ in the GHRS data. 
We will use the column density from IUE in this component.
The other values agree within the ranges of the errors.

Comparing the \ion{Mg}{ii} profile at 2803~\AA\, from IUE and GHRS spectra, 
one finds in the IUE profile what looks like a feature between $-50$ and 
$0$~km\,s$^{-1}$ (see Fig. \ref{Spectra}).
Considering the superior quality of the GHRS spectra, this feature is probably
not real.
The effect of the feature on the resulting column densities, however, is 
negligible.

The ratio $N(\mathrm{\ion{Na}{ii}})/N(\mathrm{\ion{Na}{i}})$ 
is not available from observations.
We can, however, use $N(\mathrm{\ion{Zn}{ii}})/N(\mathrm{\ion{Na}{i}})$ as a 
substitute (Vladilo \& Centuri\'on \cite{vlad_cen94}).
The only slight depletion of \ion{Zn}{ii} and its ionisation potential of
$18.0$~eV make it a good tracer of the neutral gas, 
while \ion{Na}{i} is sensitive to the physical conditions in the gas.

\begin{table}[t]
\begin{flushleft}
\caption[]{$N(\mathrm{\ion{Mg}{i}})/N(\mathrm{\ion{Mg}{ii}})$ ratios}
\label{MgI_II}
\setlength{\tabcolsep}{1.00mm}
\begin{tabular}{l*{4}{c}}
\hline\noalign{\smallskip}
      & $\log N_{\mathrm{\ion{Mg}{i}}}$ & $\log N_{\mathrm{\ion{Mg}{i}}}$ & $\log N_{\mathrm{\ion{Mg}{ii}}}$ & $\log N_{\mathrm{\ion{Mg}{i}}}/N_{\mathrm{\ion{Mg}{ii}}}$ \\
      & & (Bowen) & & \\
\noalign{\smallskip}\hline\noalign{\smallskip}
A     & --             & $12.41\pm0.02$ & $14.95\pm0.30$ & $-2.54\pm0.30$ \\
B     & --             & $12.00\pm0.08$ & $>13.90$       & $<-1.82$       \\
C     & $12.65\pm0.15$ & $12.64\pm0.06$ & $14.15\pm0.15$ & $-1.51\pm0.16$ \\
D$_a$ & $12.45\pm0.20$ & $13.63^{+0.12}_{-0.04}$ & $14.60\pm0.30$ & $<-0.55$ \\
D$_b$ & --             & $11.41\pm0.03$ & $<13.40$       & $>-2.02$       \\
E     & $11.20\pm0.35$ & $11.92\pm0.08$ & $14.70\pm0.30$ & $-3.50\pm0.46$ \\
F     & $12.95\pm0.35$ & $12.76^{+0.10}_{-0.13}$ & $14.60\pm0.30$ & $-1.84\pm0.33$ \\
G     & $11.80\pm0.15$ & $11.93\pm0.03$ & $14.20\pm0.30$ & $-2.27\pm0.30$ \\
\noalign{\smallskip}\hline
\end{tabular}
\end{flushleft}
\end{table}

\begin{table}[t!]
\begin{flushleft}
\caption[]{$N(\mathrm{\ion{Zn}{ii}})/N(\mathrm{\ion{Na}{i}})$ ratios, based 
           either on the observed \ion{Zn}{ii} and \ion{Na}{i} column densities
           or on \ion{Na}{i} alone, using Eq.~(\ref{lin_eqn})}
\label{ZnNa}
\setlength{\tabcolsep}{0.98mm}
\begin{tabular}{l*{4}{c}l}
\hline\noalign{\smallskip}
      & $\log N_{\mathrm{\ion{Zn}{ii}}}$ & $\log N_{\mathrm{\ion{Na}{i}}}$ & $\log N_{\mathrm{\ion{Na}{i}}}$ & \multicolumn{2}{c}{$\log N_{\mathrm{\ion{Zn}{ii}}}/N_{\mathrm{\ion{Na}{i}}}$} \\
\noalign{\smallskip}\cline{5-6}
 & & (Vladilo) & (Bowen) & & \\
\noalign{\smallskip}\hline\noalign{\smallskip}
A     & $12.90\pm0.20$ & $12.06$ & $12.03\pm0.01$ & $0.86\pm0.20$  & obs.  \\
B     & $12.35\pm0.25$ & --      & $<11.48$       & $>0.62$        & obs.  \\
C     & $12.05\pm0.30$ & $11.23$ & $11.34\pm0.02$ & $0.77\pm0.30$  & obs.  \\
D$_a$ & $12.95\pm0.20$ & $12.62$ & $14.09\pm0.13$ & $<0.53$        & obs.  \\
D$_b$ & --             & $12.18$ & $<10.23$       & $>-0.84$       & Eq. (\ref{lin_eqn}) \\
E     & $11.85\pm0.25$ & --      & $<10.49$       & $>1.11$        & obs.  \\
F     & $12.40\pm0.20$ & $12.89$ & $13.01\pm0.01$ & $-0.55\pm0.20$ & obs.  \\
G     & --             & --      & $<10.32$       & $>0.74$        & Eq. (\ref{lin_eqn}) \\
\noalign{\smallskip}\hline
\end{tabular}
\end{flushleft}
\end{table}

Column densities for \ion{Na}{i} are given by Vladilo et al. (\cite{vladilo94})
and by Bowen et al. (\cite{bowen94}).
We equate the higher resolved optical velocity components with our IUE 
components as defined above for \ion{Mg}{i}.
Both optical data sets are quite similar. 
Bowen et al. (\cite{bowen94}), however, have spectra of a slightly higher 
resolution than those of Vladilo et al. (\cite{vladilo94}), resulting in a 
larger number of fitted \ion{Na}{i} components.

In most cases, the \ion{Na}{i} column densities from the two optical spectra,
assigned to our component structure, are comparable (Table \ref{ZnNa}).
The values disagree for the weak component D$_b$, due to different velocity 
centres of the fitted components.
We thus take the value from Vladilo et al. (\cite{vladilo94}) as an upper 
limit.
A strong discrepancy is visible in the local component D$_a$.
The value of Vladilo et al. (\cite{vladilo94}) is assumed as a lower limit.
No \ion{Na}{i} is detected at $+160$~km\,s$^{-1}$ (component G).

In the components D$_b$ and G, where a column density for \ion{Zn}{ii} cannot 
be determined, we use instead the linear approximation of 
Vladilo \& Centuri\'on (\cite{vlad_cen94})
\begin{eqnarray}
\label{lin_eqn}
\log\frac{N(\mathrm{\ion{Zn}{ii}})}{N(\mathrm{\ion{Na}{i}})}&\simeq& -0.78(\pm0.07)\log N(\mathrm{\ion{Na}{i}})+\\
& & \hspace{3cm}+10.31(\pm0.80),\nonumber
\end{eqnarray}
based on only \ion{Na}{i}.
However, these values can only describe the general tendency and must not be
overinterpreted.

Temperature $T$ and electron density $n_e$ in a gas are related to the
ionisation ratio of an element X according to
\begin{equation}
\frac{N(\mathrm{\ion{X}{ii}})}{N(\mathrm{\ion{X}{i}})}
  =\frac{\Gamma_{\rm X}}{\alpha(T)\cdot n_e},
\end{equation}
where $\Gamma_{\rm X}$ is the photoionisation rate and $\alpha(T)$ the
recombination coefficient.
As an initial approximation for $\Gamma_{\rm X}$, we use the values WJ1 of 
de Boer et al. (\cite{deboer73}), based on the mean Galactic radiation field 
of Witt \& Johnson (\cite{witt73}).
Additional contributions from charge exchange reactions (see e.g. P\'equignot 
\& Aldrovandi \cite{pequignot86}) can be neglected, since the density of the 
gas is low.

Radiative recombination coefficients are taken from P\'equignot \& 
Aldrovandi (\cite{pequignot86}). 
For magnesium we must also take into account the dielectronic 
recombination coefficient given by Shull \& van Steenberg (\cite{shull82}), 
and the contribution of
dielectronic recombination processes at lower temperatures calculated by 
Nussbaumer \& Storey (\cite{nussbaumer86}).

\begin{figure}[t!]
\resizebox{\hsize}{!}{\includegraphics{h1837.f2}}
\caption[]{Electron densities and temperatures of the different absorption 
           components. 
           ``Mg'' denotes the plots calculated from 
           $N(\mathrm{\ion{Mg}{i}})/N(\mathrm{\ion{Mg}{ii}})$, 
           ``Na'' those from 
           $N(\mathrm{\ion{Na}{i}})/N(\mathrm{\ion{Na}{ii}})$, 
           which are derived from 
           $N(\mathrm{\ion{Na}{i}})/N(\mathrm{\ion{Zn}{ii}})$.
           From the intersepts of the curves we estimate the physical 
           parameters of the gas}
\label{IonPlots}
\end{figure}

For magnesium we therefore get a temperature dependent electron density of
\begin{equation}
n_e(T)=\frac{\Gamma_{\rm Mg}}{\alpha(T)}\cdot
       \frac{N(\mathrm{\ion{Mg}{i}})}{N(\mathrm{\ion{Mg}{ii}})},
\end{equation}
while we use for sodium
\begin{equation}
\frac{N(\mathrm{\ion{Na}{ii}})}{N(\mathrm{\ion{Na}{i}})}
=\frac{N(\mathrm{\ion{Na}{ii}})}{N(\mathrm{\ion{Zn}{ii}})}
 \cdot\frac{N(\mathrm{\ion{Zn}{ii}})}{N(\mathrm{\ion{Na}{i}})}
\end{equation}
with
\begin{equation}
\label{eqn_Na_Zn}
\frac{N(\mathrm{\ion{Na}{ii}})}{N(\mathrm{\ion{Zn}{ii}})}
\simeq \frac{N(\mathrm{Na})}{N(\mathrm{Zn})}
\simeq50,
\end{equation}
assuming solar abundances. 
The left approximation in Eq.~(\ref{eqn_Na_Zn}) is valid, because in neutral 
gas the first ionisation stages of sodium and zinc are far more abundant than 
the neutral forms.

Plotting these two functions into a diagram for each component, we can 
read the physical conditions in the gas from the line curve intersepts
(Fig. \ref{IonPlots}). 
The resulting limits for temperatures and electron densities are given in 
Table~\ref{TabTemp}.
Exact limits could not be given for all components. 

Only component C yields a well defined intersept that provides firm limits 
on $T$ and $n_e$.
For A, a value in the flat part of the diagram below $3500$~K seems to be most 
probable.
If $T$ was significantly higher in this cloud, one would expect a much 
higher state of ionisation and thus much higher electron densities in the gas,
which would not fit with the diagram.
Only D$_b$ seems to have a combination of high temperature and high 
electron density. 
For E and G we probably have temperatures between $4000$ and $10\,000$~K.

Since the curves of magnesium and sodium are nearly parallel below about
$3500$~K, we need additional information for a more precise temperature
determination.
According to Vladilo \& Centuri\'on (\cite{vlad_cen94}),
$\log R_N=\log[N(\mbox{\ion{Zn}{ii}})/N(\mbox{\ion{Na}{i}})]$ can be used 
by itself as an indicator for temperature and electron density in the gas.
In our data the column densities are not exact enough for $\log R_N$ to 
provide tight restrictions to these parameters, 
but we can at least assume cold gas around $100$~K for $\log R_N\la0.8$ and 
warm gas for $\log R_N\ga0.9$.

Under this criterion, we find that component D$_a$ and probably also F consist
of cold gas. 
While A and C are still undecided, the gas in B is probably warm.
For the gas in E and especially G we also get a tendency towards higher 
temperatures.

\begin{table}[t!]
\begin{flushleft}
\caption[]{Estimated temperatures, electron densities, hydrogen densities, 
           and line of sight extents of the identified velocity components}
\label{TabTemp}
\begin{tabular}{lcccc}
\hline\noalign{\smallskip}
  & $n_e$~[cm$^{-3}$]\,$^a$ & $T$~[K]\,$^a$ & $n_{\rm H}$~[cm$^{-3}$] & $l$ [pc] \\
\noalign{\smallskip}\hline\noalign{\smallskip}
A     & $0.02-0.2$  & $\sim100-5000$  & $\ga50\,^b$  & $2.6$    \\
B     & $<0.2$      & $3200-10000$    & $<7\,^c$     & $>1.8$   \\
C     & $0.08-0.32$ & $5000-7000$     & $2.7-11\,^c$ & $10-2$   \\
D$_a$ & $0.02-32$   & $\sim100-6000$  & $\ga50\,^b$  & $\ga1.6$ \\
D$_b$ & $<1.6$      & $\ga5000$       & $<50\,^c$    & $\la8$   \\
E     & $<0.06$     & $\sim100-10000$ & $<2\,^c$     & $>10$    \\
F     & $0.13-4.0$  & $\sim100-4000$  & $\ga300\,^b$ & $\la0.3$ \\
G     & $<0.16$     & $>3500$         & $<5\,^c$     & $\la0.6$ \\
\noalign{\smallskip}\hline\noalign{\smallskip}
\end{tabular}

$^a$ from Fig. \ref{IonPlots} \\
$^b$ adopting fully neutral H, free electrons mainly from C$^+$ \\
$^c$ adopting 3\% H$^+$ \\
\end{flushleft}
\end{table}

With known electron densities we can now roughly estimate hydrogen densities 
and hence the extents of the clouds. 
Assuming the hydrogen in the gas is totally neutral, most of the electrons 
stem from carbon, which is the most abundant ionised metal in the ISM. 
For cold gas at about $100$~K we thus can use the carbon abundance of 
$-3.4$~dex (de Boer et al. \cite{deboer87a}) to derive the hydrogen abundance.
This probably is the case in components A, D$_a$, and F.
For warm gas of several $1000$~K the fraction of ionised hydrogen can be 
$1$\% to $5$\%, depending on the level of ionising radiation in the 
environment (Vladilo \& Centuri\'on \cite{vlad_cen94}; Kulkarni \& Heiles 
\cite{kulkarni87}). 
Under this assumption, electrons from ionised hydrogen are by far dominant.
However, the wide temperature ranges derived above and the lack of information
on the radiation field in the clouds make it difficult to determine exact 
densities.
For the probably warm components B, C, D$_b$, E, and G with $T\ga 4000$~K, we 
assume a fraction of ionised hydrogen of $3$\%.

Based on the values derived for $n_{\rm H}$, 
we then calculate the line of sight extent of each 
cloud, with appropriate uncertainties.
The results are listed in Table \ref{TabTemp}.

Due to the uncertain ionisation ratio of hydrogen, especially in the warm
components, the derived densities could be smaller than 
listed in Table~\ref{TabTemp}.
McKee \& Ostriker (\cite{mckee77}) assume a fractional ionisation of 15\% 
for the warm neutral ISM, which would lead to a factor $5$ smaller 
hydrogen densities.
In the cool components A, D$_a$, and F a non-negligible fraction of ionised 
hydrogen could have an even larger effect on the hydrogen densities.

As a consequence, also the calculated cloud sizes strongly depend on the 
level of hydrogen ionisation assumed for the clouds.
The warm clouds are about a factor $5$ larger if 15\% of the hydrogen is 
ionised.
Looking at the cool components, a fractional ionisation of only 1\% already 
means a linear extent about a factor $25$ larger than stated in 
Table~\ref{TabTemp}.

%%%%%%%%%%%%%%%%%%%%%%%%%%%%%%%%%%%%%%%%%%%%%%%%%%%%%%%%%%%%%%%%%%%%%%%%%%%%%%%

\section{Higher ionisation stages towards SN\,1993J}
%%%%%%%%%%%%%%%%%%%%%%%%%%%%%%%%%%%%%%%%%%%%%%%%%%%%

\label{High_Ion}

\subsection{Column densities}
%============================

The gas seen in the absorption lines of higher ions is likely spatially 
distinct from the lower ionised gas and has to be treated separately.

For the analysis, we determine apparent column densities,
as described by, e.g., Savage \& Sembach (\cite{savage91}): 
The optical depth $\tau$ at a wavelength $\lambda$ is approximately
\begin{equation}
\tau(\lambda)\simeq\tau_a(\lambda)=\ln [I_0(\lambda)/I_{\rm{obs}}(\lambda)],
\end{equation}
with $I_0(\lambda)$ the continuum intensity and $I_{\rm{obs}}(\lambda)$ the
observed intensity at this wavelength.
For each velocity bin of $20$~km\,s$^{-1}$, which is the resolution of the IUE,
the logarithmic column density $\log[N_a(v)]$ apparently occuring in that bin
then computes as
\begin{eqnarray}
\log[N_a(v)]&=&\log\tau_a(v)-\log(f\lambda_{\rm{tr}})+15.876 \\
  & & \hspace{2cm}[\rm{atoms\,cm}^{-2}\,(20~\rm{km\,s}^{-1})^{-1}], \nonumber
\end{eqnarray}
where $f$ is the transition probability and $\lambda_{\rm{tr}}$ the wavelength
of the transition.

We examine the doublet transitions of \ion{Al}{iii} ($1854$ and $1862$~\AA), 
\ion{Si}{iv} ($1393$ and $1402$~\AA), and \ion{C}{iv} ($1548$ and $1550$~\AA). 
No absorption of the \ion{N}{v} doublet at $1238$ and $1242$~\AA\, is observed 
along this line of sight. 
This lack of \ion{N}{v} absorption was also observed towards \object{SN\,1987A}
(Fransson et al. \cite{fransson87}) and was to be expected. 
The lower solar abundance of nitrogen and the lower $f$-values for the 
\ion{N}{v} transitions lead to a nominal line strength about a factor of $6$ 
lower than that of \ion{C}{iv}, making \ion{N}{v} hard to detect in the halo. 
Furthermore, the \ion{N}{v} transitions are located at smaller wavelengths,
where the UV flux from the supernova had already decreased, thus lowering
the signal-to-noise ratio in this region of the spectra.

Reliable absorption profiles are calculated for each of the observed spectra.
The comparison of the apparent column density profiles in each doublet shows
no indication of saturation for any of the atoms, 
i.e., no unusual deviation in the peak heights.
We thus take the mean of the four available profiles 
(two transitions in each of the two spectra) of each ion.
Fig. \ref{ACD_Plots} shows the resulting mean apparent column density profiles.
\ion{C}{iv} showed an unnaturally low column density at $-40$~km\,s$^{-1}$, 
which was disregarded.

\begin{figure}[t!]
\resizebox{\hsize}{!}{\includegraphics{h1837.f3}}
\caption[]{Apparent column densities for the highly ionised atoms 
           \ion{C}{iv}, \ion{Si}{iv}, and \ion{Al}{iii}.
           The extremely low \ion{C}{iv} value at $-40$~km\,s$^{-1}$ was 
           disregarded.
           Where no column density is plotted, the profile indicated upper 
           limits lower than the end points of the profiles}
\label{ACD_Plots}
\end{figure}

Significantly larger errors for the mean apparent column densities at 
velocities lower than $-100$~km\,s$^{-1}$ (\ion{Al}{iii}: $-120$~km\,s$^{-1}$)
and larger than $+220$~km\,s$^{-1}$ (\ion{Al}{iii}: $+200$~km\,s$^{-1}$) 
indicate a lack of absorption at these velocities.
Only upper limits can be given here.
Especially near $-130$~km\,s$^{-1}$, the rotational velocity of \object{M\,81}
at the position of \object{SN\,1993J}, no absorption is found. 

We find a broad absorption component around $-40$~km\,s$^{-1}$ and another 
broad absorption between $+80$ and $+200$~km\,s$^{-1}$, the latter probably 
due to at least $2$ clouds of velocities around $+110$ and $+170$~km\,s$^{-1}$.
Also, a weak component at $0$~km\,s$^{-1}$ is visible. 
The integrated column densities for these components are listed in Table
\ref{High_CDs93}.

\begin{table}[t!]
\begin{flushleft}
\caption[]{High ion metal column densities towards \object{SN\,1993J}. 
           Only upper limits can be given for gas at $-130$~km\,s$^{-1}$.
           The errors for the other values are in the range of $0.15$ dex.
           In the rightmost column the total column density for absorption
           above $+80$~km\,s$^{-1}$ is listed}
\label{High_CDs93}
\setlength{\tabcolsep}{1.19mm}
\begin{tabular}{l*{6}{c}}
\hline\noalign{\smallskip}
 & \multicolumn{6}{c}{$\log \left[\sum_v N(v)\right]$~[cm$^{-2}$]} \\
\noalign{\smallskip}\cline{2-7}\noalign{\smallskip}
$v_{\rm{LSR}}$~[km\,s$^{-1}$] & $-130$ & $-40$ & $0$ & $+110$ & $+170$ & $110\,\&\,170$ \\
\noalign{\smallskip}\hline\noalign{\smallskip}
\ion{C}{iv}   & $<13.1$ & $14.3$ & $\simeq13.0$ & $14.0$ & $13.9$ & $14.3$ \\
\ion{Si}{iv}  & $<12.8$ & $14.0$ & $\simeq12.8$ & $13.7$ & $13.7$ & $14.0$ \\
\ion{Al}{iii} & $<12.3$ & $13.1$ & -- & $12.9$ & $12.9$ & $13.2$ \\
\noalign{\smallskip}\hline
\end{tabular}
\end{flushleft}
\end{table}

\subsection{Physical conditions in the gas}
%==========================================

\label{HighConditions}

Ionisation of gas to the level of \ion{Si}{iv} and \ion{Al}{iii} can be 
explained by stellar photons. 
For \ion{C}{iv} the necessary ionisation energy is too large to be 
provided by stellar radiation,
but these ions can be easily produced, for example, in the cooling process 
of a galactic fountain.

Looking at the apparent column density profiles, we find similar profiles
for \ion{Si}{iv} and \ion{C}{iv} in the negative velocity range, with a nearly
constant ratio $\log[N(\rm{\ion{C}{iv}})/N(\rm{\ion{Si}{iv}})]$ of about $2.5$.
The profile of \ion{Al}{iii} differs, but this might be due to the weakness
of the absorption.
Both \ion{C}{iv} and \ion{Si}{iv}
possibly exist in the same spatial environment, in a border region 
between hot, infalling fountain gas and cooler, photoionised or recombined,
gas.
This spatial coexistence seems to be a common phenomenon in halo gas, it was 
also found by Savage et al. (\cite{savage89}) on the line of sight to
\object{SN\,1987A}.

Nothing comparable is found in the positive velocity range.
The profile of \ion{C}{iv} clearly differs from that of \ion{Si}{iv}. 
We probably have spatially distinct clouds at similar velocities here, 
possibly in the vicinity of \object{M\,81}.

%%%%%%%%%%%%%%%%%%%%%%%%%%%%%%%%%%%%%%%%%%%%%%%%%%%%%%%%%%%%%%%%%%%%%%%%%%%%%%%

\section{Spatial arrangement of the clouds}
%%%%%%%%%%%%%%%%%%%%%%%%%%%%%%%%%%%%%%%%%%%

\label{SN93}
\label{sub_space_highion}

An estimate of the locations of the absorbing clouds is difficult for the 
highly ionised gas. 
However, we can distinguish three basic groups of gas: 
gas within the Milky Way, either in the disk or in its halo, gas in 
\object{M\,81}, and intergalactic gas at some position in the space between 
\object{M\,81} and the Galaxy, probably near the \object{M\,81} group.

\subsection{Gas in the Milky Way}
%================================

In the low ion transitions, components C, D$_a$, and D$_b$ are considered to
be Galactic gas.
Their velocities and their clear visibility in the $21$~cm radio emission
spectra let any other interpretation than nearby gas seem improbable.

{\bf Component C}, from its velocity of $-50$~km\,s$^{-1}$, is by definition
an IVC in our Galaxy, probably part of the Low Latitude Intermediate Velocity 
(LLIV) Arch (Kuntz \& Danly \cite{kuntz96}).
The large $b$-value and column density indicate a possible blend of more than 
one cloud.

{\bf Component D$_a$} shows the typical parameters for cold local disk gas.

{\bf Component D$_b$} may represent absorption by gas of the Local Cloud,
in which our solar system is imbedded. 
For Local Cloud gas in the direction of \object{M\,81}, G\'enova et al. 
(\cite{genova90}) find a temperature around $11500$~K and a hydrogen density 
of $n_H\simeq0.1$~cm$^{-3}$ at a velocity of $+11$~km\,s$^{-1}$.
The magnesium column density has a low value of only 
$\log N(\mathrm{Mg})<12.1$.
The extent of this gas is about $3-4$~pc.
These parameters agree nicely with our results.

In the profiles of the highly ionised atoms we find additional 
Milky Way absorption components:
The absorption around $0$~km\,s$^{-1}$ has its origin probably in Galactic
disk gas, possibly in gas of the Local Bubble, surrounding the Local Cloud.
The gas seen around $-40$~km\,s$^{-1}$ could be related to the IVC component 
C in neutral gas. 
A transition between neutral and ionised gas also would explain the 
coexistence of \ion{C}{iv} and \ion{Si}{iv} found in Chapter 
\ref{HighConditions}.
We obviously have no contribution from the nearby Galactic HVC Complex C, 
which, in this general direction,
shows velocities of $-100$~km\,s$^{-1}$ and below (Wakker \cite{wakker91}). 

\subsection{Gas in M\,81}
%========================

We find neutral gas in absorption at velocities near $-130$~km\,s$^{-1}$. 
From the radiointerferometry map of \object{M\,81} gas by Rots 
(\cite{rots75}), this is the rotational velocity of the galaxy at the position
of the supernova.
Peak-shaped emission is not visible in the \ion{H}{i} 21~cm emission spectra at
this velocity, but rather only broad background emission from the disk of 
\object{M\,81}.
The gas in absorption is thus probably located at a large distance, at which 
it could not be detected separately in the $9\arcmin$ radio beam 
of the Effelsberg telescope.
For these reasons the components A and B can be assumed to be located in the 
disk or the lower halo of \object{M\,81}.

In {\bf component A}, the \ion{H}{i} column density is comparable to local gas.
The metal column densities in this component are very similar to those in
Milky Way gas.
We probably see cold \object{M\,81} disk gas at $-130$~km\,s$^{-1}$ here.

Also assuming a location in \object{M\,81}, {\bf component B} can be 
interpreted as an IVC or a blend of such clouds, at a velocity of 
$-40$~km\,s$^{-1}$ relative to the \object{M\,81} disk.

No highly ionised gas is detected that would have to be associated with a 
location near \object{M\,81}.

\subsection{Intergalactic gas in front of M\,81}
%===============================================

The three absorption components E, F, and G seen at positive velocities 
can hardly be assigned to the Galaxy or \object{M\,81} itself. 
They must be intergalactic clouds, located probably in the IGM of the 
\object{M\,81} group. 
The gas seen in the highly ionised atoms at positive velocities is 
associated with intergalactic clouds as well.

The hydrogen column density of the cooler {\bf component E} is comparable to 
that of the IVC components B and C, with metal depletions in a normal range.

In the cold {\bf component F} we find a hydrogen column density about one 
order of magnitude smaller than for the local Galactic component D$_a$. 
The depletion is low, with presumably no depletion at all for oxygen. 
This is an indicator for a low dust content in the gas. 
If we assume a location in the vicinity of \object{M\,81}, the interpretation 
as an infalling \object{M\,81} HVC is possible.
The low temperature, however, contradicts an origin from a galactic 
fountain mechanism. 

{\bf Component G} shows warm
gas of very low hydrogen column density and low metal depletion.
Similar to component E, it could be high velocity gas still in its cooling 
phase, infalling towards \object{M\,81}.

\vspace*{0.2cm}

We find very low electron densities for the warm neutral gas components E and 
G at positive velocities.
This is unusual, since our calculations are based on the Galactic disk 
radiation field, which is too large for halo gas.
The disk radiation fields of the Galaxy and \object{M\,81} should be 
comparable.
However, if the components are located at some distance from the disk of 
\object{M\,81}, as assumed from their velocities, the level of radiation should
be significantly lower, leading to even lower electron densities
than derived above. 
We thus conclude that an enhanced level of ionising radiation exists at the 
location of the components E and G, at least at the level of a disk radiation 
field.
Evidence for enhanced ionising radiation in a range typical for the Galactic
disk was also inferred by Vladilo et al. 
(\cite{vladilo94}) from the $N(\mathrm{\ion{Ca}{ii}})/N(\mathrm{\ion{Ca}{i}})$
ratio.

Radiation from hot disk stars is an improbable source of energy for this 
level of ionisation. 
The interpretation of typical HVCs seems unlikely under these conditions.
If located in the foreground of \object{M\,81}, the dynamic intergalactic 
medium of the \object{M\,81} group itself could provide the energy needed. 
The source of energy is unclear, but could be shocks or plasma effects
like magnetic reconnection (Birk et al. \cite{birk98}).
Also mixing effects at the border region between hot intergalactic gas and
cooler tidal debris are possible.

The low pressure in the IGM would explain the probably large extent of the hot
component E.
Component F, from its velocity assigned to this intergalactic gas, is possibly
the cold, shielded core of some intergalactic cloud.
This would explain the high hydrogen density at a low extent.
The high density $n_{\mathrm{H}}\ga300$~cm$^{-3}$ derived assuming free 
electrons solely from carbon might indicate that at least a small fraction of 
hydrogen is ionised in this component.
This would mean a lower density and a larger extent, rather typical for an
extended cold cloud.

The gas seen in the high ion absorption lines at velocities larger than 
$+80$~km\,s$^{-1}$ must also be located in front of \object{M\,81}. 
Bregman (\cite{bregman80}) predicts mainly negative velocities in Galactic 
neutral gas in the direction of \object{M\,81}.
Of course, the highly ionised gas cannot be identical to this neutral gas.
Nevertheless, if both ionisation stages are only different states of infalling
gas, their velocities should be comparable. 
Positive velocities around $+140$~km\,s$^{-1}$, however, have been found by 
Yun et al. (\cite{yun93}) in the tidal tail region connecting \object{M\,81}
and \object{M\,82}.

Since we find indicators for a high energy environment from the lower 
ionisation stages, the coexistence of highly and low ionised gas and cold, 
shielded cloud cores in the \object{M\,81} group IGM is plausible.
Similar conditions were recently found by Sembach et al. (\cite{sembach00}) 
in the Magellanic Stream and in outer Galaxy HVCs, 
where neutral \ion{H}{i} gas is visible together with \ion{O}{vi} at the
interface to the surrounding hot medium.

%%%%%%%%%%%%%%%%%%%%%%%%%%%%%%%%%%%%%%%%%%%%%%%%%%%%%%%%%%%%%%%%%%%%%%%%%%%%%%%

\section{Combined SN\,1987A to SN\,1993J line of sight}
%%%%%%%%%%%%%%%%%%%%%%%%%%%%%%%%%%%%%%%%%%%%%%%%%%%%%%%

\label{Long_LoS}

\begin{figure}[t!]
\resizebox{\hsize}{!}{\includegraphics{h1837.f4}}
\caption[]{Schematical view of the lines of sight from the Milky Way to the 
           two nearby supernovae. 
           In the lower right the Magellanic Clouds are visible at about 
           50~kpc distance; \object{M\,81} at 3.63~Mpc is found in the upper 
           left corner}
\label{Pic_Long_LoS}
\end{figure}

\subsection{Line of sight and velocity structure}
%================================================

The antipodal directions of the two supernovae allow us to combine their lines
of sight into a single one that traverses three galaxies in a direction nearly
perpendicular to the plane of the Local Group (Fig. \ref{Pic_Long_LoS}).
As noted by de Boer et al. (\cite{deboer87b}), the position of 
\object{SN\,1987A} is on the backside of the \object{LMC}.
Starting there, the line of sight crosses the local \object{LMC} gas, the HVCs and IVCs of the 
southern Galactic halo, the full disk of the Milky Way, and its northern halo
with the IVC. 
It continues on, sampling the intergalactic medium towards the \object{M\,81} 
group and the halo of \object{M\,81}, before it reaches the disk of 
\object{M\,81} itself at the location of \object{SN\,1993J}.
The total length of this line of sight is about $3.7$~Mpc, sampling gas at
a total velocity width of $500$~km\,s$^{-1}$.

To create a spectrum over the combined line of sight we merge the absorption 
line spectra of the two supernovae.
We examine the transitions of \ion{Si}{ii} at $1304$, $1526$ and $1808$~\AA, as
well as the \ion{C}{iv} transition at $1548$~\AA.
For the combination we use the ripple corrected spectra and sum over the 
absorption in each velocity bin.
We do lose some equivalent width in those absorption components which are 
saturated. It is shown below (Sect.~\ref{CombinedCD}), however, that this 
loss has a negligible effect on the total equivalent width.
It is not neccessary for this examination to add the optical depths and 
generate Voigt profiles of the absorption, which is the mathematically correct
procedure.
Fig.~\ref{Comb_Abs} shows the combined absorption profiles for the total line
of sight.
We see a complex blending of absorptions from the different galaxies in the 
absorption profile. 
Without knowledge about the individual galaxies it is impossible to associate 
individual absorption components with a particular galaxy.

\begin{figure}[t]
\resizebox{\hsize}{!}{\includegraphics{h1837.f5}}
\caption[]{Absorption profiles created for the combined line of sight from
           \object{SN\,1987A} to \object{SN\,1993J}, 
           based on the spectra to each of the supernovae.                     
           \object{SN\,1987A} to \object{SN\,1993J}. 
           The velocity origin is set to the position of
           \object{SN\,1987A}. 
           The marks in the top of the figure indicate which absorption 
           structure is due to which of the individual gas components along 
           the total line of sight. 
           They now heavily blend, of course}
\label{Comb_Abs}
\end{figure}

\subsection{Column densities}
%============================

\label{CombinedCD}

We obtain the equivalent widths for the complete absorption structure by a 
trapezium fit.
For \ion{Si}{ii} we can fit these values to a theoretical curve of growth with
a doppler parameter of $b\simeq80$~km\,s$^{-1}$. 
A $b$-value that large was to be expected for such massive blending of 
absorption components. 
The total absorption leads to a column density of 
$\log N_{\mathrm{total}}(\mathrm{\ion{Si}{ii}})=16.1$.
If we just sum over the equivalent widths of the single components and fit
these to a curve of growth, we get the same value of 
$\log N_{\mathrm{total}}(\mathrm{\ion{Si}{ii}})=16.1$, but with a $b$-value of 
$100$~km\,s$^{-1}$.
Obviously, the column density for these data is not very sensitive to changes 
in the equivalent width. 
This justifies the procedure of just adding the absorption profiles instead
of correctly adding the optical depths.

From the sum over the single column densities we get
$\log N_{\mathrm{total}}(\mathrm{\ion{Si}{ii}})=16.6$, which is about a factor
of $3$ larger than from the curve of growth.
This deviation shows that, in absorption profiles with heavy line blending, 
the total column density can easily be underestimated by a significant amount.

No reliable total column density can be determined for \ion{C}{iv}. 
With only one data point we cannot use the curve of growth method.
The apparent column density method is not suitable here, due to the saturation
of the profile.
The sum over the single column densities is
$\log N_{\mathrm{total}}(\mathrm{\ion{C}{iv}})\ga15.0$, using the value of 
$14.7$ (Savage et al. \cite{savage89}) for \object{SN\,1987A} and the values 
from Table \ref{High_CDs93} for \object{SN\,1993J}.

%%%%%%%%%%%%%%%%%%%%%%%%%%%%%%%%%%%%%%%%%%%%%%%%%%%%%%%%%%%%%%%%%%%%%%%%%%%%%%%

\section{Comparison with quasar lines of sight}
%%%%%%%%%%%%%%%%%%%%%%%%%%%%%%%%%%%%%%%%%%%%%%%

\label{QSO_LoS}

Complex absorption structures seen in the spectra of quasars are commonly 
identified with absorption by galaxies, galactic halos,
or other diffuse objects such as dwarf galaxies along the line of sight 
(see, e.g., Savage \& Jeske \cite{savage81b}, York et al. \cite{york86},
Yanny \& York \cite{yanny92}).
We now compare typical quasar absorption with the absorption structure 
found on our Local Group line of sight.
We use equivalent widths listed for three different quasar sight lines from
Savage \& Jeske (\cite{savage81b}) and those of one quasar in the Hubble Deep
Field South from Savaglio (\cite{savaglio98}).
The velocity widths of the absorption complexes are between $150$ and
$550$~km\,s$^{-1}$, which is comparable to the $500$~km\,s$^{-1}$ found for 
the local absorption. 
The data are listed in Table \ref{Quasar_CDs}.

\begin{table}[t!]
\begin{flushleft}
\caption[]{Quasar column densities in comparison to column densities on the
           line of sight from \object{SN\,1987A} to \object{SN\,1993J}}
\label{Quasar_CDs}
\setlength{\tabcolsep}{1.5mm}
\begin{tabular}{lcrrrc}
\hline\noalign{\smallskip}
 & $z$  & \multicolumn{3}{c}{$\log N$ [cm$^{-2}$]} & Note\,$^a$ \\
\noalign{\smallskip}\cline{3-5}\noalign{\smallskip}
 &      & \multicolumn{1}{c}{\ion{Si}{ii}} & \multicolumn{1}{c}{\ion{C}{iv}} & \multicolumn{1}{c}{$\frac{\mathrm{\ion{C}{iv}}}{\mathrm{\ion{Si}{ii}}}$} & \\
\noalign{\smallskip}\hline\noalign{\smallskip}
\object{QSO 1756$+$237}  & 1.673 & $>14.3$\hspace{1.85mm} & $\la15.8$\hspace{1.9mm} & $<1.5$    & a \\
\object{QSO 0002$-$422}  & 2.302 & 14.9\hspace{1.85mm}    & 14.9\hspace{1.9mm}      & 0.0       & b \\
\object{PKS 2126$-$158}  & 2.638 & 14.8\hspace{1.85mm}    & $>15.0$\hspace{1.9mm}   & $>0.2$    & c \\
                         & 2.769 & 14.3\hspace{1.85mm}    & 14.7\hspace{1.9mm}      & 0.4       & c \\
\object{QSO J2233$-$606} & 1.787 & $<14.9$\hspace{1.85mm} & 14.3\hspace{1.9mm}      & $>-0.6$   & d \\
                         & 1.869 & $<14.8$\hspace{1.85mm} & 14.5\hspace{1.9mm}      & $>-0.3$   & d \\
                         & 1.928 & $<13.6$\hspace{1.85mm} & $<13.7$\hspace{1.9mm}   & $\sim0.1$ & d \\
                         & 1.943 & $<14.0$\hspace{1.85mm} & $<14.8$\hspace{1.9mm}   & $\sim0.8$ & d \\
                         & 2.077 & $<13.6$\hspace{1.85mm} & $13.3$\hspace{1.9mm}    & $>-0.3$   & d \\
                         & 2.198 & $<13.3$\hspace{1.85mm} & 13.8\hspace{1.9mm}      & $>0.5$    & d \\
                         & 2.206 & $<14.3$\hspace{1.85mm} & 14.2\hspace{1.9mm}      & $>-0.1$   & d \\
\noalign{\smallskip}\hline\noalign{\smallskip}
\multicolumn{2}{l}{\object{SN\,1987A} to \object{SN\,1993J}}    & $>16.1$\hspace{1.85mm} & $\ga15.0$\hspace{1.9mm} & $\la-1.1$   & \\
\multicolumn{2}{l}{\object{LMC} gas}& $16.4\,^b$ & $>14.8\,^c$ & $>-1.6$ & \\
\multicolumn{2}{l}{Clouds F+G to \object{SN\,1993J}}    & $14.5$\hspace{1.85mm}  & $14.3$\hspace{1.9mm}    & $-0.2$    & \\
\noalign{\smallskip}\hline\noalign{\smallskip}
\end{tabular}
$^a$ Data from (a) Turnshek et al. (\cite{turnshek79});
               (b) Sargent et al. (\cite{sargent79}); 
               (c) Young et al. (\cite{young79}); 
               (d) Savaglio (\cite{savaglio98})\\
$^b$ Gas at $v_{\rm{LSR}}>190$~km\,s$^{-1}$, from Welty et al. (\cite{welty99})\\
$^c$ Gas at $v_{\rm{LSR}}>190$~km\,s$^{-1}$, from Savage et al. (\cite{savage89})
\end{flushleft}
\end{table}

The column densities in \ion{Si}{ii} are generally lower on the quasar sight 
lines than in the \object{SN\,1987A} - \object{SN\,1993J} gas, while in 
\ion{C}{iv} the values are similar.
The relation of lower to higher ionisation stages is thus significantly 
lower on the quasar sight lines.
Both results are expected, since the local line of sight is an extreme case
of three galactic disks along one line of sight. 
Most of the neutral gas is found in the disks of galaxies, so this explains 
the high column density in \ion{Si}{ii}.
The halos of galaxies are far more extended than the disks, so the probability
of sampling halo gas in a quasar sight line is far higher than for disk gas.
This leads to the observed relation of lower to higher ionisation stages
in the quasar absorption complexes and to the generally lower column density
in the less dense halo gas.

Dwarf galaxy gas or gas of any other diffuse nature may also produce 
absorption lines. 
Substantial \ion{Si}{ii} may be present in dwarf galaxies, as in 
the case of the \object{LMC}. 
Free gas outside galaxies containing metals is most likely material expelled 
by such galaxies.

The ratio $N($\ion{C}{iv}$)/N($\ion{Si}{ii}$)$ could possibly be used as
an indicator for halo gas along lines of sight. 
For local gas we get a value of $<-1.1$~dex, while the values for quasar 
absorption complexes range between $-0.6$ and $1.5$~dex.
This ratio can be influenced by different depletions of the elements, however,
especially since both ionisation stages exist in completely different 
environments. 

In particular, we can compare the cloud complex F+G located in the 
intergalactic medium in front of \object{M\,81} with these results.
Since we do not detect high ions at the velocity of \object{M\,81} component E,
we consider only the ratio for components F and G alone 
(see Table \ref{Quasar_CDs}). 
In contrast to the total line of sight, the ratio 
of $N($\ion{C}{iv}$)/N($\ion{Si}{ii}$)$ for the intergalactic cloud complex 
is very similar to those from QSO absorption systems, with comparably low 
column densities.

%%%%%%%%%%%%%%%%%%%%%%%%%%%%%%%%%%%%%%%%%%%%%%%%%%%%%%%%%%%%%%%%%%%%%%%%%%%%%%%

\section{The intergalactic clouds in front of M\,81}
%%%%%%%%%%%%%%%%%%%%%%%%%%%%%%%%%%%%%%%%%%%%%%%%%%%%

\label{IGCs}

The cloud complex F+G found in the foreground of \object{M\,81} does not show 
a low density, metal poor structure 
as one usually assumes for intergalactic gas.
Instead, it shows a mixture of cold and warm, normal density, metal rich gas,
in coexistence with highly ionised hot components,
as normally expected on lines of sight crossing a galactic disk.
This indicates that the F+G complex probably contains gas of galactic disk 
origin.

The complex extragalactic gas features of the \object{M\,81} group
have been mapped in emission of both \ion{H}{i} (see, e.g., Cottrell 
\cite{cottrell77}; Appleton et al. \cite{appleton81};
Appleton \& van der Hulst \cite{appleton88}) and CO (e.g., 
Brouillet et al. \cite{brouillet91}).
Identifying gas as lying in front of the bright \object{M\,81} disk, however,
is difficult from emission.
The existence of a separate cloud would follow from the difference
of its velocity with respect to the velocity of the gas of the background 
object (here \object{M\,81}), either in \ion{H}{i} or in CO.
The UV absorption lines show that the detected gas is indeed in front of 
\object{M\,81}.
Furthermore, they provide us with full information on metallicities and 
ionisation ratios in the intergalactic gas in this foreground region.

Molecular gas was found in the intergalactic medium surrounding \object{M\,81}
(Brouillet et al. \cite{brouillet92}; Walter \& Heithausen \cite{walter99}).
We do not know if molecules exist in component F+G, since no radio molecules 
were looked for and the visual spectra of \object{SN\,1993J} gave only upper 
limits for CH$^+$ (Vladilo et al. \cite{vladilo94}).
Given the low temperature of absorption component F (see 
Sect.~\ref{Phys_Cond}), the existence of molecular gas in the F+G cloud 
complex is possible.

The origin of the gas on our line of sight is possibly \object{M\,82}.
The absorption velocities are comparable to the ones found by Yun et al. 
(\cite{yun93}) for the \ion{H}{i} tidal tail connecting \object{M\,81} and 
\object{M\,82}.                                             
A recent simulation by Sofue (\cite{sofue98}) indicates that \object{M\,82}
could have lost its complete disk in an encounter with \object{M\,81}.
If this were so, we would expect processed disk gas in the foreground of 
\object{M\,81}, possibly also containing stars.
The Sofue scenario could thus explain the metallicities as well as the 
galaxy disk-like ionisation structure of the gas.
We note here, that another stripping event may have taken place in the 
\object{M\,81} group.
Walter \& Heithausen (\cite{walter99}) found CO gas displaced with respect to 
the optically well-defined galaxy \object{NGC\,3077}, 
i.e., here without indications for associated stars.

Our procedure to derive electron densities in the gas components is based on 
the average Milky Way disk radiation field.
The level of radiation should be different in intergalactic space far away 
from galaxy disks. 
Nevertheless, the intergalactic absorption components show signs for 
ionisation comparable to that in the radiation field of the Galaxy or even 
larger. 
Another explanation for the observed level of ionisation could be the presence
of a stellar component in the gas, 
not unrealistic in a scenario of material ripped from \object{M\,82} 
in a close encounter.
If \object{M\,82} indeed lost its spiral arms, 
stars could well be present in the intergalactic medium of the \object{M\,81} 
group.                                                          

The detection of highly ionised gas in the complex F+G demonstrates 
that neutral and ionised gas are mixed, very similar to what is seen in 
QSO absorption line systems.
It is unclear, however, just where exactly the ionised gas is situated.
It is either in the border region of the hot intergalactic plasma, 
thermally excited by collisions, or it is gas excited by a radiation source 
somewhere within the \object{M\,81} system.

%%%%%%%%%%%%%%%%%%%%%%%%%%%%%%%%%%%%%%%%%%%%%%%%%%%%%%%%%%%%%%%%%%%%%%%%%%%%%%%

\begin{acknowledgements}
%%%%%%%%%%%%%%%%%%%%%%%%

OM is supported by Deutsche Forschungsgemeinschaft grant Bo~779/24.
The data used in this paper were provided by the ESA IUE Final Archive at 
VILSPA, Spain.
We thank the referee for comments which helped to improve the paper.

\end{acknowledgements}

%%%%%%%%%%%%%%%%%%%%%%%%%%%%%%%%%%%%%%%%%%%%%%%%%%%%%%%%%%%%%%%%%%%%%%%%%%%%%%%

\end{document}